\documentclass[aps,pra,superscriptaddress,10pt]{revtex4-2}
\usepackage{amsthm}
\usepackage{amsmath}
\usepackage{amssymb}
\usepackage{color}
\usepackage{braket}
\usepackage{url}
\usepackage{mathptmx}
\usepackage{txfonts}
\usepackage{graphicx}

\usepackage[colorlinks,citecolor=magenta,linkcolor=red]{hyperref}
\usepackage[usenames,dvipsnames,svgnames]{xcolor}

\begin{document}

\title{ Macroscopically distinguishable superposition in infinitely many degrees of freedom }
\author{Jonas Fransson}
\affiliation{Department of Physics and Astronomy, Materials Theory, Uppsala University, 751 37 Uppsala, Sweden}
\email{jonas.fransson@physics.uu.se}
\author{Barry C.\ Sanders}
\affiliation{Institute for Quantum Science and Technology,
University of Calgary,
Alberta T2N 1N4, Canada}
\email{sandersb@ucalgary.ca}
\author{Artur P. Sowa}
\affiliation{Department of Mathematics and Statistics, University of Saskatchewan, Saskatchewan S7N 5E6, Canada}
\email{sowa@math.usask.ca}
\date{\today}

\maketitle

\textbf{Abstract: }
    We investigate the concept of macroscopically distinguishable superpositions within an infinite array of boson sites.  Our approach is rigorous within the frame of Hilbert space theory.
In this context, it is natural to differentiate between states—and corresponding dynamics—that involve only finitely many degrees of freedom, referred to as local, and those that are inherently nonlocal. Previous studies have shown that such systems can support nonlocal coherent states (NCS). In this work, we demonstrate that NCS can dynamically evolve into nonlocal cat states under the influence of a nonlocal Hamiltonian—specifically, the square of the total number operator. Crucially, the resulting dynamics cannot be decomposed into local factors.
Furthermore, we explore broader mathematical implications of these phenomena within the framework of generalized bosons. Our findings highlight that the concepts of coherent states and nonlocal cat states are not inherently bound together; rather, their fusion is a distinctive feature of standard bosons. Finally, we propose that if the generalized boson framework can be physically realized in engineered quantum systems, the phenomena described here may hold significant relevance for both physics and materials science.  

\section{Introduction}

The concept of a Schr\"{o}dinger cat state—a particular type of quantum superposition—is of critical importance as a hallmark of quantumness. The experimental realization of such states serves as a key validation of quantum mechanics. One of the enduring puzzles in the field is why these distinctly nonclassical phenomena fail to manifest at macroscopic scales. Consequently, the creation of macroscopically distinguishable cat states remains central to efforts aimed at deepening our understanding of quantum theory. In recent years, significant progress has been made in this direction, with experimental observations of cat states reported in systems ranging from atomic nuclei~\cite{GVY+24} to nanogram-scale mechanical structures~\cite{BFY+23}.
In turn,
Girvin surveys the development of cat-state experiments within the framework of circuit quantum electrodynamics, including a discussion of their connection to bosonic quantum error correction codes~\cite{Gir19}.

Contemporary investigations build on foundational theoretical work, notably the seminal papers~\cite{Mil86} and~\cite{YS86}, which analyzed the phenomenon of creation of cat states—wherein a coherent state (CS) evolves under a nonharmonic Hamiltonian into a superposition of two macroscopically distinct coherent states. In those works, the essential mechanism for producing cat states involves the unitary evolution of a coherent state under the nonharmonic Hamiltonian $\hat{N}^2$, the square of the number operator.  

In this work, we revisit the cat-state creation phenomenon in a new setting, replacing the standard CS with a nonlocal coherent state (NCS)—a coherent state defined over an infinite array of bosonic sites. Here, \emph{nonlocality} has a purely spatial meaning---extended support over infinitely many sites---and is not meant to invoke quantum nonlocality. The NCS were introduced in~\cite{SF25} in a rigorous framework. Their construction involves the indexing of local sites by consecutive prime numbers, rather than integers—a technique first used in~\cite{Spe90} and~\cite{BC95}. This leads to a natural identification of the bosonic Fock space of the entire array with the Hilbert space
\[
\ell_2(\mathbb{N}) = \mbox{span}\,\, \{ \ket{n}: \,\, n \in \mathbb{N}\}.
\] 
Note that $\ket{1}$ is the vacuum state, while for a prime $p$, states of the form $\ket{p^k}$, $k\in \mathbb{N}$, are interpreted as being localized at site $p$, for a pair of primes $p,q$ states of the form $\ket{p^kq^l}$, $k, l\in \mathbb{N}$, are localized on two sites, etc. In this way, every Fock basis state $\ket{n}$, $n>1$, is localized on finitely many sites, as indicated by the prime decomposition of the number $n$. 
The key idea behind the construction of NCS is to use an Euler-type formula to define an infinite tensor product of local CS rigorously.  While the notation deployed here is completely equivalent to the standard one (see Subsection \ref{Section_NT}), it is indispensable for the formulation and manipulation of such structures.  Recall, the classical Euler product has the form:
\[
\prod\limits_{p} (1+ p^{-s} + p^{-2s} + p^{-3s} + \ldots ) = \sum_{n=1}^{\infty} n^{-s},
\]
where the product on the left is over the set of all prime numbers, and $s$ is a complex variable or a fixed complex number, depending on the application.   
Replacing the infinite product of numbers with an infinite symmetric tensor product enables one to exploit this pattern in the context of an infinite array of boson sites. Indeed, some Fock states may be viewed as products of localized states, e.g.,
\[
\bigodot\limits_{p} \left(\ket{1} + \frac{p^{-s}}{\sqrt{1!}} \ket{p} + \frac{p^{-2s}}{\sqrt{2!}} \ket{p^2}
+  \frac{p^{-3s}}{\sqrt{3!}} \ket{p^3} + \ldots \right)
= \sum\limits_{n}  \frac{n^{-s}}{c_n}\, |\,n\,\rangle  ,
\]
where $c_n$ are suitably defined constants. Note that this is a Fock state, i.e. an element of $\ell_2(\mathbb{N})$, whenever $\Re s> 1/2$.
The NCS arise via a construction of this type, as discussed in detail in Subsection \ref{subsection_NCS_construction}.  In particular, the NCS are given as a superposition of all Fock basis states, and are therefore, indeed, nonlocal. 
The NCS are eigenvectors of an infinite family of annihilation operators $\hat{a}_p$ where, again, $p$ is an arbitrary prime. Correspondingly, there are infinitely many local number operators $\hat{N}_p$, and a global number operator $\hat{N} = \sum_p \hat{N}_p$. This gives rise to two distinct nonharmonic Hamiltonians: the locally quadratic Hamiltonian $\mathcal{H} = \sum_p \hat{N}_p^2$, and the globally quadratic Hamiltonian $\mathcal{H}_g = \hat{N}^2$. As we will demonstrate, these two cases exhibit fundamentally different dynamical behavior.

To summarize:

\begin{itemize}
  \item The evolution of the NCS induced by $\mathcal{H}$ is separable. Viewed through the Q-function formalism, one observes the same type of evolution at every site—akin to the regular case (see~\eqref{QtNCSnonh})—and the complete Q-function is a product of local Q-functions, each evolving independently.

  \item In contrast, the evolution under $\mathcal{H}_g$ is not separable (see~\eqref{Q_NCS_global}). Instead, it is essentially equivalent to the regular case, but arises through a construction that involves all local parameters simultaneously (see~\eqref{glob_loc_subst}).

  \item The cat-state creation phenomenon does not manifest in the NCS case when the evolution is based on the locally quadratic Hamiltonian $\mathcal{H}$. A superficial version of this effect may still appear when the Hamiltonian is nonharmonic at only finitely many sites, but this occurs for trivial reasons (see~\eqref{sz_doubling2}).

  \item The cat-state creation is clearly manifest when the NCS evolves under the globally quadratic Hamiltonian $\mathcal{H}_g$ (see~\eqref{sz_doubling_glob}). Moreover, the Yurke-Stoler interference patterns are fully consistent with those observed in the regular case and can be independently detected at any single site (see Section~I).
\end{itemize}  

The mathematical structures used to demonstrate the cat-state creation in the context of nonlocal coherent states (NCS) closely resemble those found in prior literature. In Section \ref{Section_gen_bosons}, we move beyond this framework by introducing a specific model of generalized bosons and reexamining both the notion of coherent states and the cat-state creation within this new context. Notably, we find that these two concepts, which coincide in the standard theory, now diverge: the coherent states are no longer subject to cat-state creation under a nonharmonic Hamiltonian, yet a distinct set of states does exhibit this behavior. From this perspective, the unification of these notions in the conventional bosonic setting appears as a miraculous mathematical coincidence.

Separately, the question of whether the specific generalized bosons introduced here can manifest physically remains open. However, it is plausible that such systems could be engineered or discovered. Indeed, certain types of generalized bosons are known to be physical (see, e.g.,~\cite{KXH+22} and references therein).

\section{Types of dynamics that generate the nonlocal cat-state creation }          

In this section, we examine the types of dynamics that either preclude or ensure an analogue of the cat-state creation phenomenon in the context of nonlocal coherent states. A brief review of the local picture sets the stage and helps to introduce notation and terminology.

\subsection{The regular coherent state in the homodyne detector variable}
Recall, see e.g.~\cite{KS85}, that the regular coherent states, denoted $|\, \alpha\rangle$ with $\alpha\in \mathbb{C}$, have the form
\begin{equation}\label{alpha}
\ket{\alpha} = e^{-|\alpha|^2/2}\sum_{k=0}^{\infty} \, \frac{\alpha^{k}}{\sqrt{k!}} \, \, | k\rangle,
\end{equation}
and satisfy 
\begin{equation}\label{eig_alpha}
  \hat{a} \ket{\alpha} = \alpha \, \ket{\alpha}. 
\end{equation}
The homodyne operator is defined as
\begin{equation}\label{hatx}
 \hat{x} = \left(1/\sqrt2\right)\, \left[e^{i\theta} \hat{a}+e^{-i\theta} \hat{a}^\dagger\right],
\end{equation}
where the local-oscillator phase $\theta \in [0, 2\pi)$ is controlled by the experimenter. The term `homodyne operator' follows the tradition of~\cite{YS86}. It is specific to optics, where $\hat{x}$ represents a homodyne measurement. The phase of the homodyne measurement allows tunability for measuring an arbitrary linear combination of position and momentum. However, variants of this terminology may be more common in other types of physics applications. 

We are interested in expressing the coherent state in the variable corresponding to $\hat{x}$. That variable is defined via the relation  
\begin{equation}\label{xprime}
  \hat{x} \, \ket{x'} = x'\, \ket{x'}.
\end{equation}
To be sure, the state $\ket{x'}$ does not exist (as an element of the Hilbert space), it is only used virtually. Indeed, when $\theta =0$, $\hat{x}$ is multiplication by $x$ so, formally $\ket{x'} = \delta_{x'}$. Nevertheless, this approach can be made rigorous on the grounds of nuclear spectral theory. Alternatively, in the special case, it can be approached via the Wigner-Weyl transform and rotations in the $(x,p)$ plane. We will handle this special case directly as follows:
\vspace{.5cm}

We need to calculate $\psi_\alpha (x'): = \langle x' | \alpha \rangle$. To this end,
apply $\langle x'|$ to 
\begin{equation}
\label{psi1}
  \alpha  \, \psi_\alpha (x')   = \langle x' | \,  \hat{a} \, \ket{\alpha}. 
\end{equation}
Next, introduce a new operator:
\begin{equation}\label{hatp}
 \hat{p} = \left(1/\sqrt2\right)\, \left[-i e^{i\theta} \hat{a}+i e^{-i\theta} \hat{a}^\dagger\right],
\end{equation}
Note that 
\begin{equation}\label{commut}
  [\, \hat{x}, \hat{p}\,] = i.
\end{equation}
Note that $\hat{x}$ acts on a state that is represented in variable $x$ by multiplication by $x$. Therefore, $\hat{p}$ will act on such a state in the same representation as $-i d/dx$. This is the basic principle underlying the calculation that follows. Indeed, we observe that
\begin{equation}\label{aequals}
  \hat{a} = \frac{e^{-i\theta}}{\sqrt{2}} \left( \hat{x} + i \hat{p}\right).
\end{equation} 
It follows that 
\begin{equation}\label{calc}
  \langle x' | \,  \hat{a} \, \ket{\alpha} = \frac{e^{-i\theta}}{\sqrt{2}} \,\langle x' | \, \hat{x} + i \hat{p} \, \ket{\alpha}
  =  \frac{e^{-i\theta}}{\sqrt{2}} \, \left( x'\, \, \psi_\alpha (x') + \frac{d}{dx'} \, \psi_\alpha (x')\right).
\end{equation}
The last identity follows from the fact that applying an operator to a state and then expressing the result in the variable $x'$ is equivalent to applying the same operator (in adapted form) to that state after it has been expressed in variable $x'$. 
  Combining this with (\ref{psi1}) we obtain an ODE:
  \begin{equation}\label{ODE}
    \frac{d}{dx'} \, \log \psi_\alpha (x') =  - x' + \sqrt{2} e^{i\theta} \alpha.
  \end{equation}
  Solving (\ref{ODE}) by integration (with $x'\mapsto x$) gives:
   \begin{equation}\label{psifinal}
     \psi_\alpha (x) =  \frac{1}{\pi^{1/4}} \exp\left( -\frac{x^2}{2} + \sqrt{2} e^{i\theta}\alpha x - \left(\frac{e^{i\theta}\alpha + e^{-i\theta}\alpha^*}{2} \right)^2\right).
   \end{equation} 
   The choice of the constant of integration, namely, $- (\Re{e^{i\theta}\alpha})^2$, ensures that $\int_{\mathbb{R}} |\psi_\alpha (x)|^2 dx = 1$. In many physics articles and books, including~\cite{YS86}, the constant is chosen as  $- ((e^{i\theta}\alpha)^2 + |\alpha|^2)/2$, which leads to $L_2$ norm different than $1$. 

\subsection{Bosons on an infinite array of sites: a multiplicative notation}
\label{Section_NT}

It is indispensable for us to use the multiplicative notation in the construction of bosonic Fock space, first introduced in~\cite{Spe90} and~\cite{BC95}.
Namely, we fix prime numbering of the standard basis in the single-particle Hilbert space, i.e., $\mathbb{H}_{\text{SP}} = \text{span}\{|p\rangle:\, p \text{ prime}\}$. By definition, the many-particle Fock space has the following structure:
\[
\mathbb{H}^\odot = \bigoplus\limits_{k=0}^\infty \mathbb{H}_{\text{SP}}^{\odot k}, \quad \mbox{ where }  \, \, \mathbb{H}_{\text{SP}}^{\odot 0} = \mathbb{C}.
\]
Here, $\odot$ signifies the symmetric tensor product. Note that the subspace $\mathbb{H}_{\text{SP}}^{\odot k}$ is spanned by vectors of the form
$
|p_1\rangle \odot\cdots\odot |p_k\rangle$,
where $p_1, \ldots p_k$ is any collection of $k$ primes, possibly with repetitions. Uniqueness of the prime decomposition of integers allows one to identify
\[
|p_1\rangle\odot  \cdots\odot |p_k\rangle = | n \rangle,
\]
where $n = p_1 \cdot \ldots \cdot p_k$ is the prime decomposition of $n$. Thus, $\mathbb{H}^\odot = \mbox{ span }\{ |n\rangle: n \in \mathbb{N}\} = \ell_2(\mathbb{N}).$  Next, we define $a_p(n)$ as the multiplicity of $p$ in the prime decomposition of $n$, so that
\[
 n = \prod_{p} \, p^{a_p(n)} .
\]
Note that only finitely many terms in the product are different than $1$. For every prime $p$, i.e., $p = 2,3,5,\ldots$, the annihilation and creation operators, $\hat{a}_p$ and $\hat{a}_p^\dagger$, are then defined via their effect on basis vectors:
\begin{equation}\label{as}
    \hat{a}_p \, \ket{ n} = \sqrt{a_p(n)}\,\, \ket{n/p },  \quad
     \hat{a}_p^\dagger \, \ket{n}  = \sqrt{a_p(n)+1}\,\, \ket{ np}.
\end{equation}
It is easily seen that the bosonic Canonical Commutation Relations are satisfied, i.e.
\begin{equation}\label{BCCR}
[\,  \hat{a}_p, \hat{a}_q^\dagger \, ] = \delta_{p,q}, \quad  [\,  \hat{a}_p, \hat{a}_q \, ] = 0, \quad  [\,  \hat{a}_p^\dagger, \hat{a}_q^\dagger \, ] = 0\quad \mbox{ for all primes } p,q.
\end{equation}
We also have the local number operators $\hat{N}_p = \hat{a}_p^\dagger \hat{a}_p$ and the total (or global) number operator $\hat{N} = \sum_{p} \hat{N}_p$. Note that $\hat{N} \, \ket{n} = \Omega (n) \ket{n}$, where $\Omega(n)$ is the big prime omega function, i.e., the number of prime divisors of $n$ counted with multiplicities. Furthermore, we have the following characterization 
\begin{equation}\label{k-part}
  \mathbb{H}_{\text{SP}}^{\odot k} = \text{span}\{\ket{n}:\,  \Omega(n) = k\}.
\end{equation}
  The following table illustrates the equivalence between the multiplicative notation and the standard one. 
\vspace{.8cm} 

\begin{tabular}{|l|r|r|}
  \hline
   & the multiplicative notation & a transcript to the standard notation \\
   \hline 
   &&\\
  a basis vector  & $\quad \ket{n}\quad$ & $\ket{ a_2(n), a_3(n), \ldots, a_p(n), \ldots}\quad $ \\
  \hline
  && \\
    annihilation $\hat{a}_p$ on a basis vector& $\sqrt{a_p(n)}\,\, \ket{n/p }\quad$  &  $\sqrt{a_p(n)}\,\, \ket{ a_2(n), a_3(n), \ldots, a_p(n)-1, \ldots}\quad $  \\
   \hline 
   &&\\
    creation $\hat{a}_p^\dagger$ ... & $ \sqrt{a_p(n)+1}\,\, \ket{ pn}\quad$  & $\sqrt{a_p(n) +1 }\,\, \ket{ a_2(n), a_3(n), \ldots, a_p(n) +1, \ldots}\quad $ \\
  \hline
   & & \\
   local number operator $\hat{N}_p$ ... & $ a_p(n) \ket{n}  \quad$ & $a_p(n)\,\, \ket{ a_2(n), a_3(n), \ldots, a_p(n), \ldots}\quad $\\
   \hline
\end{tabular}

\subsection{The case of Nonlocal Coherent States (NCS)}\label{subsection_NCS_construction}

\begin{figure}[t]
\centering
\includegraphics[width=90mm]{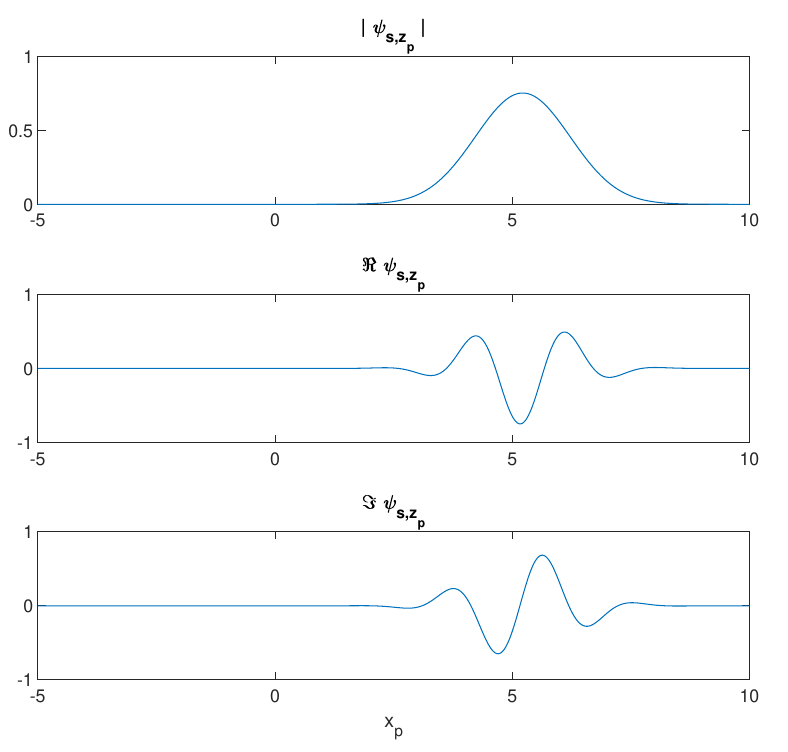}
\caption{An example of $\psi_{p^{-s}z_p} (x_p)$ given by (\ref{psip}). Here: $p = 2, s = 1+1i, z_p = 3+8i, \theta_p = \pi/3$. The Gaussian $|\psi_{p^{-s}z_p}|^2$ is centered around $x_p \simeq 5.2178$, and has variance $1/\sqrt{2}$. }
\label{Numer}
\end{figure}

The NCS have been introduced in~\cite{SF25}. Here, we give only a basic description of the NCS that is needed to serve the main purposes of this article.   
To this end, we focus on the most general form of the NCS, which enables the resolution of the identity formula. First, we define a sequence of constants $c_n$ given by 
\begin{equation}\label{magic_cs}
    c_n = \sqrt{\prod\limits_p a_p(n)!} 
\end{equation}
Second, we utilize a sequence of complex variables, labeled by primes, namely $\vec{z} = (z_2, z_3, z_5, \ldots )\in \mathbb{C}^\omega$. Finally, we select a complex parameter $s = \sigma +it$, where $\sigma>1/2$ and $t$ is an arbitrary real number. With this understood, we define the NCS via:
\begin{equation}\label{sz}
  \ket{ s, \vec{z}\, }   = e^{-\frac{1}{2}   P(2\sigma, |\vec{z}|^2)} \sum_{n=1}^\infty \frac{n^{-s}}{c_n}\, \prod z_p^{a_p(n)}\, |\,n\,\rangle,
  \quad \mbox{ where }\quad
 P(2\sigma, |\vec{z}|^2) = \sum_{p} p^{-2\sigma}|z_p|^2.
\end{equation}
 The normalization factor imposes a constraint on $\vec{z} = (z_2, z_3, z_5, \ldots )\in \mathbb{C}^\omega$, namely $ P(2\sigma, |\vec{z}|^2) <\infty$. This ensures that $|\, s, \vec{z}\, \rangle$ is a vector in the Fock space. The main observation is that $|\, s, \vec{z}\, \rangle$ can be represented as an infinite symmetric tensor product. In fact, this is foundational and enabling for the construction. It explains why there is a need for the Dirichlet series in the construction.  And so:
\begin{equation}\label{inf_tens_pr}
|\, s, \vec{z}\, \rangle =  \bigodot\limits_{p} \, \ket{p^{-s}z_p}, \quad  \mbox{ where }\,\,  \ket{p^{-s}z_p}= \exp(-p^{-2\sigma} |z_p|^2/2) \,\sum_{k=0}^{\infty} \, \frac{(p^{-s}z_p)^k}{\sqrt{k!}} \, \, |p^k\rangle .
\end{equation}
 Note that 
\begin{equation}\label{appn}
    \hat{a}_p \, | p^k \rangle = \sqrt{k}\,\, |p^{k-1}\rangle,  \quad
     \hat{a}_p^\dagger \, |p^k\rangle  = \sqrt{k+1}\,\, | p^{k+1}\rangle,
\end{equation}
and each $\ket{p^{-s}z_p}$ is a local (one-site) coherent state, i.e., 
\begin{equation}\label{apszp}
 \hat{a}_p \, \ket{p^{-s}z_p} = p^{-s}z_p \, \ket{p^{-s}z_p}.
\end{equation}
This also determines the action of $\hat{a}_p$ on the NCS, namely:
\begin{equation}\label{apsvecz}
 \hat{a}_p \, \ket{s, \vec{z}} = p^{-s}z_p \, \ket{s, \vec{z}}.
\end{equation}
Substitution $p^{-s}z_p \mapsto \alpha_p$ and $| p^k \rangle \mapsto | k \rangle$ allows one to interpret each $\ket{p^{-s}z_p}$  as a regular coherent state. 
We now have a family of homodyne detectors, indexed by primes, namely:
\begin{equation}\label{hatxp}
 \hat{x}_p = \left(1/\sqrt2\right)\, [\, e^{i\theta_p} \hat{a}_p + e^{-i\theta_p} \hat{a}_p^\dagger \, ].
\end{equation}
Due to separation of variables given by (\ref{inf_tens_pr}), $ \hat{x}_p$ acts on $\ket{p^{-s}z_p}$ (and is extended by identity to other factors). Setting $\alpha = \alpha_p = p^{-s}z_p$ and $\theta = \theta_p$ in (\ref{psifinal}), we obtain: 
   \begin{equation}\label{psip}
     \psi_{p^{-s}z_p} (x_p) = \frac{1}{\pi^{1/4}} \exp\left( -\frac{x_p^2}{2} + \sqrt{2} e^{i\theta_p}p^{-s}z_p  x_p - \left(\Re(e^{i\theta_p}p^{-s}z_p )\right)^2\right).
   \end{equation}
Note that the center of the Gaussian $|  \psi_{p^{-s}z_p}|^2$ is at $x_p = \sqrt{2}\Re(e^{i\theta_p}p^{-s}z_p )$ and the variance is $1/\sqrt{2}$, see Fig.\ref{Numer}. It is interesting to observe that due to the convergence condition in (\ref{sz}), the sequence $|\Re(e^{i\theta_p}p^{-s}z_p )|\leq |p^{-s}z_p |$, where $p = 2,3, 5,\ldots $ needs to be square-summable. Therefore, the centers of consecutive Gaussians  $|  \psi_{p^{-s}z_p} (x_p)|$  will converge to $0$. However, any finite choice of them may have arbitrary position on their respective $x_p$ axes. 

Importantly, the confluence of (\ref{apszp}) and (\ref{apsvecz}) indicates that the NCS, when ``seen" in the homodyne variable $x_p$ is identical with the local CS, namely:
   \begin{equation}\label{psipcvecz}
     \psi_{s, \vec{z}} (x_p) = \psi_{p^{-s}z_p} (x_p) .
   \end{equation}

\subsection{The Q-function}

We will adapt the discussion in~\cite{Mil86} to the NCS. To this end, we begin by a review of the single-site basics. For an evolving mixed state $\rho=\rho(t)$, the Q-function is defined via
\begin{equation}\label{Q-def}
  Q(\alpha, \alpha^*, t) = \langle \alpha |\, \rho(t)\, \ket{\alpha}, \mbox{ so that } \quad \pi^{-1}\int d^2\alpha \, Q(\alpha, \alpha^*, t) = 1.
\end{equation}
We are interested in the particular types of dynamics and the initial condition $\rho(0) = |\alpha_0 \rangle\langle \alpha_0 |$. In particular,
\begin{equation}\label{Q0}
   Q(\alpha, \alpha^*, 0) = e^{-|\alpha - \alpha_0|^2}.
\end{equation}
Suppose at first, the evolution is governed by the Hamiltonian $\mathcal{H} = \hat{N} = \hat{a}^\dagger \hat{a}$, so that $\rho(t) = e^{- i \hat{N}t} \rho(0) e^{i\hat{N}t}$. We have  
\[
e^{ i \hat{N}t}  \ket{\alpha} = e^{-|\alpha|^2/2}\sum_{k=0}^{\infty} \, \frac{\alpha^{k}}{\sqrt{k!}} \, \, e^{i \hat{N}t}\, | k\rangle =e^{-|\alpha|^2/2}\sum_{k=0}^{\infty} \, \frac{\alpha^{k}}{\sqrt{k!}} \, \, e^{ i kt}\, | k\rangle = |e^{it}\alpha\rangle,
\]
which implies
\begin{equation}\label{Qt}
   Q(\alpha, \alpha^*, t) = e^{-|e^{it}\alpha - \alpha_0|^2} = e^{-|\alpha - e^{-it}\alpha_0|^2}  \quad \mbox{(harmonic)}.
\end{equation}
Thus, as $t$ increases, the centre of the Gaussian evolves circularly clockwise. 

We are interested in the evolution of the Q-function when the Hamiltonian is nonharmonic (alternatively, a nonharmonic correction to the harmonic part, in which case one passes to the moving frame picture). Specifically, we take $\mathcal{H} = \hat{N}^2$, so that $\rho(t) = e^{- i \hat{N}^2t} \rho(0) e^{i\hat{N}^2t}$. We have  
\begin{equation}\label{alpha_evol_nhm}
 e^{ i \hat{N}^2t}  \ket{\alpha} = e^{-|\alpha|^2/2}\sum_{k=0}^{\infty} \, \frac{\alpha^{k}}{\sqrt{k!}} \, \, e^{ i \hat{N}^2t} \, | k\rangle =e^{-|\alpha|^2/2}\sum_{k=0}^{\infty} \, \frac{\alpha^{k}}{\sqrt{k!}} \, \, e^{ i k^2t} \, | k\rangle.
\end{equation}
This is not a canonical coherent state anymore. A direct calculation gives
\begin{equation}\label{Qtnonh}
   Q(\alpha, \alpha^*, t) = e^{-|\alpha|^2 - |\alpha_0|^2} |S(t)|^2, \mbox{ where }\quad  S(t) = \sum_{k=0}^{\infty} \frac{(\alpha_0^*\alpha)^k}{k!} e^{ik^2t}  \quad \mbox{( nonharmonic)}.
\end{equation}
No doubt, $S(t)$ is a very interesting function.

\subsection{The Q-function for the NCS}

We extend the evolution of the Q-function to the case of NCS. The definition of $Q$ is analogous to (\ref{Q-def}), except now the mixed state $\rho$ is acting in the Fock space, and so
\begin{equation}\label{Q_NCS_def}
  Q(\vec{z}, \vec{z^*}, t) = \langle s, \vec{z} \, |\, \rho(t)\, \ket{s,\vec{z}}. 
\end{equation}
  Here, $s = \sigma + i\tau$, with $\sigma > 1/2$, is arbitrary but fixed throughout. Clearly, $Q$ is nonnegative. Also, using the Hilbert-Schmidt norms, we have
  \[
  \langle s, \vec{z} \, |\, \rho(t)\, \ket{s,\vec{z}} = \mbox{ Tr } \left( \rho(t)\, \ket{s,\vec{z}}\langle s, \vec{z} \, |\right) \leq \mbox{ Tr } \left( \rho(t)^2 \right)^{1/2}
  \mbox{ Tr } \left(  \ket{s,\vec{z}}\langle s, \vec{z} \, |\right)^{1/2} \leq 1.
  \]
  Thus, $Q$ is bounded above by $1$. Furthermore, we have a resolution of the identity for the NCS. To invoke it, we need the polar coordinates, $z_p = r_p \exp (2\pi i \mu_p)$ and, consequently, $ \ket{s,\vec{z}} = |s, \vec{r}, \vec{\mu}\rangle$. With this understood, we have 
  \begin{equation}\label{res_identity}
  \int_{\mathbb{R}_+^\omega}\, d\vec{\chi} \, e^{ P(2\sigma, |\vec{r}|^2)} \,\int_{\mathbb{T}^\omega}\,d\vec{\mu} \,\, |\, s, r, \vec{\mu} \, \rangle \langle \, s, r, \vec{\mu}\, | = \sum_{k} \, \ket{k}\langle\, k\, | = I,
\end{equation}
where $ d\chi_p = d  \exp(-p^{-2\sigma}r_p^2) = 2 \exp(-p^{-2\sigma}r_p^2) p^{-2\sigma} r_p \, dr_p$, so that
\begin{equation}\label{d_chi_p}
  \int_{0}^{\infty}\, d\chi_p = 1.
\end{equation}
The measures used here, $d\vec{\mu}$ and $d\vec{\chi}$, are both Borel measures with respect to the Tychonoff topology; for details, see~\cite{SF25}.

The resolution of the identity (\ref{res_identity}) implies that
\begin{equation}\label{Q=prob}
  \mbox{ Tr } \rho = \int_{\mathbb{R}_+^\omega}\, d\vec{\chi} \, e^{ P(2\sigma, |\vec{r}|^2)} \,\int_{\mathbb{T}^\omega}\,d\vec{\mu} \,\,  \mbox{ Tr }\left( |\, s, r, \vec{\mu} \, \rangle \langle \, s, r, \vec{\mu}\,|\, \rho \right)
  =  \int_{\mathbb{R}_+^\omega}\, d\vec{\chi} \, e^{ P(2\sigma, |\vec{r}|^2)} \,\int_{\mathbb{T}^\omega}\,d\vec{\mu} \,\,
    Q(\vec{z}, \vec{z^*}, t) = 1.
\end{equation}
In other words, the $Q$ function is a probability distribution with respect to the Borel measure
\begin{equation}\label{Bmeas}
  e^{ P(2\sigma, |\vec{r}|^2)} \, d\vec{\chi} \,d\vec{\mu} . 
\end{equation}

\subsection{Evolving the NCS by the harmonic and nonharmonic Hamiltonians}  
   In particular, the initial condition is now $\rho(0) = |s, \vec{z}_0\rangle\langle s, \vec{z}_0|$. Taking into account (\ref{inf_tens_pr}) we readily obtain an analogue of (\ref{Q0}), namely:
\begin{equation}\label{Q0NCS}
   Q(\vec{z}, \vec{z^*}, 0)  = \prod\limits_p Q_p(z_p, z_p^*,0)= \exp\left(-\sum_{p} p^{-2\sigma}|z_p - z_{p,0}|^2\right).
\end{equation}
For the harmonic Hamiltonian $\mathcal{H} = \sum_{p} \hat{N}_p$, we have
\begin{equation}\label{inf_tens_prNCS}
e^{i\mathcal{H}t} \, \ket{ s, \vec{z}} =  \bigodot\limits_{p} \,e^{i\hat{N}_pt} \, \ket{p^{-s}z_p} =
\bigodot\limits_{p} \, \ket{ e^{it}p^{-s}z_p}.
\end{equation}
This leads to 
\begin{equation}\label{QtNCS}
   Q(\vec{z}, \vec{z^*}, t) = \prod\limits_p Q_p(z_p, z_p^*,t) = \prod\limits_p e^{-p^{-2\sigma}|z_p - e^{-it}z_{p,0}|^2} = 
   \exp\left(-\sum_{p} p^{-2\sigma}|z_p - e^{-it}z_{p,0}|^2\right) \quad \mbox{(harmonic)}.
\end{equation}
This is just a product of rotating Gaussians. 

Next, we consider the case of the nonharmonic Hamiltonian 
\begin{equation}\label{nonh1}
   \mathcal{H} = \sum_{p} \hat{N}_p^2,\quad \mbox{ so that } \quad \mathcal{H} \ket{n} = Q(n) \ket{n},
\end{equation}
where $Q(k) = \sum_{p} a_p(k)^2$.
Utilizing (\ref{inf_tens_pr}) again, we obtain
\begin{equation}\label{EvolNCSnonh}
  \begin{split}
     &e^{i\mathcal{H}t} \, |\, s, \vec{z}\, \rangle =  \bigodot\limits_{p} \,e^{i\hat{N}_p^2t}\, \ket{p^{-s}z_p}  \\
       &  \\
      &e^{i\hat{N}_p^2t} \, \ket{p^{-s}z_p} = e^{-p^{-2\sigma} |z_p|^2/2} \,\sum_{k=0}^{\infty} \, \frac{(p^{-s}z_p)^k}{\sqrt{k!}} e^{ik^2t}\, \, |p^k\rangle .
  \end{split}
\end{equation}
It follows that 
\begin{equation}\label{QtNCSnonh}
  \begin{split}
     & Q(\vec{z}, \vec{z^*}, t) = \prod\limits_p Q_p(z_p, z_p^*,t) \\
       &  \\
      &Q_p(z_p, z_p^*,t) =  e^{-p^{-2\sigma} (|z_p|^2 + |z_{p,0}|^2)} |S_p(t)|^2, \mbox{ where }\quad  S_p(t) = \sum_{k=0}^{\infty} \frac{\left(p^{-2\sigma}z_{p,0}^*z_p\right)^k}{k!} e^{ik^2t}  \quad \mbox{(locally quadratic)}
  \end{split}
\end{equation}
In summary, the evolution of the Q-function is separable; it is indistinguishable from the regular CS case at each local site of the array.   
It is also interesting to carry out the product over all primes, which gives
\begin{equation}\label{Q_NCS_total}
\begin{split}
    &  Q(\vec{z}, \vec{z^*}, t)= e^{-\sum_{p} p^{-2\sigma} (|z_p|^2 + |z_{p,0}|^2)}\,\, |S(t)|^2 \\
     &  \\
     & S(t) = \sum_{n=1}^{\infty} \frac{1}{c_n^2} \prod\limits_p  \left(p^{-2\sigma}z_{p,0}^*z_p\right)^{a_p(n)} e^{iQ(n)t} \quad \mbox{(locally quadratic)}\\
     & \\
     & c_n^2 = \prod\limits_p a_p(n)!, \quad Q(n) = \sum_{p} a_p(n)^2.
\end{split}
 \end{equation}
We observe that $S(t)$ is well-defined, because  its defining series converges absolutely. Indeed,
\begin{align}
\label{welldef1}
\sum_{n=1}^{\infty} \frac{1}{c_n^2}&
\prod\limits_p  \left|\left(p^{-2\sigma}z_{p,0}^*z_p\right)^{a_p(n)} e^{iQ(n)t} \right| 
=\prod_{p} \sum_{k=0}^{\infty} \frac{\left(p^{-2\sigma}|z_{p,0}^*z_p|\right)^k}{k!}
    \nonumber\\
=&\prod_{p} \exp\left( p^{-2\sigma}|z_{p,0}^*z_p|\right) = \exp \left(\sum_{p} p^{-2\sigma}|z_{p,0}^*z_p| \right) < \infty,
\end{align}
where the last inequality follows from the assumption (\ref{sz}). 

\subsection{The macroscopically distinguishable superposition}

The nonharmonic evolution (\ref{alpha_evol_nhm}) is discussed in detail in~\cite{YS86}.   Note that
\begin{equation}\label{eiksq}
  e^{ik^2\pi/2}= \frac{1}{\sqrt{2}}\left( e^{i\pi/4} +(-1)^k e^{-i\pi/4}\right) = \left\{\begin{array}{cc}
                    1, & \mbox{ for } k = 0,2,4,\ldots \\
                    i, & \mbox{ for } k = 1,3,5,\ldots 
                  \end{array}\right. 
\end{equation}
Let $|\alpha, t\rangle =  e^{ i \hat{N}^2t}  \ket{\alpha}$. (Note: In order to be consistent with the calculations presented in previous sections, we will propagate states via $\exp{ i \mathcal{H} t}$, rather than the more customary  $\exp{ -i \mathcal{H} t}$. This is of no consequence to physical conclusions.) It follows from the above and from (\ref{alpha_evol_nhm}) that
\begin{equation}\label{alpha_doubling}
  |\alpha, t+ \pi/2\rangle = \frac{e^{i\pi/4}}{\sqrt{2}} \, \ket{\alpha,t} + \frac{e^{-i\pi/4}}{\sqrt{2}} \, \ket{-\alpha,t}. 
\end{equation}
The NCS analogue of (\ref{alpha_evol_nhm}), based on the Hamiltonian $\mathcal{H} = \sum_{p} \hat{N}_p^2$, follows from (\ref{EvolNCSnonh}). Namely, for $|\, s, \vec{z}, t \, \rangle  = e^{i\mathcal{H}t} \, |\, s, \vec{z}\, \rangle $, and 
$|\, s, z_p, t \, \rangle  = e^{i\hat{N}_p^2 t} \, \ket{p^{-s}z_p} $, we have
\begin{equation}\label{EvolNCSpi2}
     |\, s, \vec{z}, t + \pi/2\, \rangle =   \bigodot\limits_{p} \,|\, p^{-s}z_p, t+ \pi/2\, \rangle =
       \bigodot\limits_{p} \, e^{-p^{-2\sigma} |z_p|^2/2} \,\sum_{k=0}^{\infty} \, \frac{(p^{-s}z_p)^k}{\sqrt{k!}} e^{ik^2t}\, e^{ik^2\pi/2} \, |p^k\rangle .
\end{equation}
Each of the local factors may be represented in the form
\begin{equation}\label{Doubling_local}
  \ket{p^{-s}z_p, t+ \pi/2} = \frac{e^{i\pi/4}}{\sqrt{2}} \, \ket{p^{-s} z_p, t} + \frac{e^{-i\pi/4}}{\sqrt{2}} \, \ket{ -p^{-s}z_p, t}.
\end{equation}
Distributing the product (\ref{EvolNCSpi2}), we obtain:
\begin{equation}\label{NCSflow_nhm}
 |\, s, \vec{z}, t+\pi/2\, \rangle = e^{-\frac{1}{2} \sum_p p^{-2\sigma} |z_p|^2} \,\sum_{n=1}^{\infty} \, \frac{n^{-s}}{c_n} \prod_{p} z_p^{a_p(n)} e^{i Q(n) t}\, e^{i Q(n) \pi/2} \, \ket{n}, \quad \mbox{ where } \quad Q(n) = \sum_{p} a_p(n)^2 .
\end{equation}
 Next, note that (\ref{eiksq}) implies
\begin{equation}\label{eiksqforQ}
  e^{i Q(n)\pi/2}= \prod\limits_{p} e^{ia_p(n)^2\pi/2} = \prod\limits_{p}  \frac{1}{\sqrt{2}}\left( e^{i\pi/4} +(-1)^{a_p(n)} e^{-i\pi/4}\right) = i^{\,\omega_2(n)},
  \end{equation}
  where $\omega_2(n)$ is the number of odd $a_p(n)$. Thus, formula (\ref{NCSflow_nhm}) yields
  \begin{equation}\label{NCSflow_nhm2}
 |\, s, \vec{z}, t +\pi/2\, \rangle  = e^{-\frac{1}{2} \sum_p p^{-2\sigma} |z_p|^2} \,\sum_{n=1}^{\infty} \, \frac{n^{-s}}{c_n} \prod_{p} z_p^{a_p(n)}  e^{i Q(n) t}\, i^{\,\omega_2(n)}\, \ket{n}.
\end{equation}
\vspace{.5cm}

\noindent
\emph{Remarks.} Proceeding naively, one might attempt to view the nonlocal state $|\, s, \vec{z}, t + \pi/2\, \rangle$ as having the form:
\[
``\,\,\bigodot\limits_p \left(\frac{e^{i\pi/4}}{\sqrt{2}} \, \ket{p^{-s}z_p, t} + \frac{e^{-i\pi/4}}{\sqrt{2}} \, \ket{ -p^{-s}z_p, t}\right)\,\, " 
\]
However, this expression is meaningless as, indeed, expanding the tensor product into a sum, one would obtain an infinite power of $1/\sqrt{2}$ as the amplitude of every term.   

Note also that local actions, via a finite sum of $\hat{N}_p^2$, result in partial factorizations involving only finitely many factors. For example, when $\mathcal{H} = \sum_{p<x}\hat{N}_p^2$, we have 
\begin{equation}\label{sz_doubling2}
|s, \vec{z}, t+ \pi/2\rangle = \bigodot\limits_{p<x} \left(\frac{e^{i\pi/4}}{\sqrt{2}} \, \ket{p^{-s}z_p, t} + \frac{e^{-i\pi/4}}{\sqrt{2}} \, \ket{-p^{-s}z_p, t}\right)\bigodot\limits_{p\geq x} \ket{p^{-s}z_p, t}.
\end{equation}
This is a superposition of finitely many discernible NCS. 
 Alternatively, a similar state can be achieved with an infinite Hamiltonian such as $\mathcal{H} = \sum_{p< x} \hat{N}_p^2 + \sum_{p\geq x} \hat{N}_p$, namely:
\begin{equation}\label{sz_doubling3}
|s, \vec{z}, t+ \pi/2\rangle = \bigodot\limits_{p<x} \left(\frac{e^{i\pi/4}}{\sqrt{2}} \, \ket{p^{-s}z_p, t} + \frac{e^{-i\pi/4}}{\sqrt{2}} \, \ket{-p^{-s}z_p, t}\right)\bigodot\limits_{p\geq x} |e^{it}p^{-s}z_p, t\rangle.
\end{equation}

\subsection{The NCS cat-state creation via evolution with a globally quadratic Hamiltonian}\label{subsection_doubling_NCS}

A closer NCS analogue of (\ref{alpha_doubling}) is obtained by selecting a nonharmonic Hamiltonian as the square of the total number operator $\hat{N}= \sum_{p} \hat{N}_p$. Thus, 
\begin{equation}\label{nonh2}
    \mathcal{H}_g = \hat{N}^2= \left(\sum_{p} \hat{N}_p\right)^2, \quad \mbox{ so that }\quad  \mathcal{H}_g \ket{n} = \Omega(n)^2 \ket{n},
\end{equation}
where, as usual, $\Omega(n) = \sum_{p} a_p(n)$ is the number of prime divisors of $n$ counting with multiplicities. Note the distinction from the Hamiltonian given in (\ref{nonh1}). It seems appropriate to view $\mathcal{H}_g$ as a globally quadratic Hamiltonian, as it accounts for products $\hat{N}_p\hat{N}_q$ for all pairs of primes $p,q$. 
In this case, we have
\begin{equation}\label{NCSflow_nhm_glob}
 \ket{s, \vec{z}, t} : =
 e^{i\mathcal{H}_g t} \, |\, s, \vec{z}\, \rangle = e^{-\frac{1}{2} \sum_p p^{-2\sigma} |z_p|^2} \,\sum_{n=1}^{\infty} \, \frac{n^{-s}}{c_n} \prod_{p} z_p^{a_p(n)} e^{i\, \Omega(n)^2 t}\, \, \ket{n}.
\end{equation}
We apply identity (\ref{eiksq}) to yield
\begin{equation}\label{eiksq2}
  e^{i\Omega(n)^2\pi/2}= \frac{1}{\sqrt{2}}\left( e^{i\pi/4} +(-1)^{\Omega(n)} e^{-i\pi/4}\right).
  \end{equation}
Note also that 
\begin{equation}\label{products}
  \prod_{p} z_p^{a_p(n)} (-1)^{\Omega(n)} = \prod_{p} (-z_p)^{a_p(n)}.
\end{equation}   
Therefore, (\ref{NCSflow_nhm_glob}) yields
\begin{equation}\label{sz_doubling_glob}
\ket{s, \vec{z}, t+ \pi/2} = \frac{e^{i\pi/4}}{\sqrt{2}} \, \ket{s, \vec{z}, t } + \frac{e^{-i\pi/4}}{\sqrt{2}} \, \ket{s, -\vec{z}, t}. 
\end{equation}
This formula is closely analogous to (\ref{alpha_doubling}). It is a manifestation of the Milburn-Yurke-Stoler cat-state creation phenomenon in the case of infinitely many boson sites. 

Analogously as in (\ref{Q_NCS_total}), the Q function is in this case calculated to be

\begin{equation}\label{Q_NCS_global}
\begin{split}
    &  Q(\vec{z}, \vec{z^*}, t)= e^{-\sum_{p} p^{-2\sigma} (|z_p|^2 + |z_{p,0}|^2)}\,\, |S(t)|^2 \\
     &  \\
     & S(t) = \sum_{n=1}^{\infty} \frac{1}{c_n^2} \prod\limits_p  \left(p^{-2\sigma}z_{p,0}^*z_p\right)^{a_p(n)} e^{i\Omega(n)^2 t} \quad \mbox{(globally quadratic)}\\
     & \\
     & c_n^2 = \prod\limits_p a_p(n)!, \quad \Omega(n)^2 = \left(\sum_{p} a_p(n)\right)^2.
\end{split}
 \end{equation}
The same reasoning as in (\ref{welldef1}) shows that $S(t)$ is well-defined in this case also. In fact, this function is of the same type as the classical, i.e., single-site, $S(t)$ given in (\ref{Qtnonh}). Indeed, a calculation shows that
\begin{equation}
S(t) = \sum_{n=1}^{\infty} \frac{1}{c_n^2} \prod\limits_p  \left(p^{-2\sigma}z_{p,0}^*z_p\right)^{a_p(n)} e^{i\Omega(n)^2 t} 
= \sum\limits_{k = 0}^\infty \frac{a^k}{k!} e^{i k^2t},\quad  \mbox{ where }\
a = \sum_p p^{-2\sigma} z_{p,0}^*z_p.
\end{equation}
In summary, there is no essential difference between this case and the regular one captured in (\ref{Qtnonh}). Indeed, the evolution is tantamount to the classical one via a nonlocal substitution:
\begin{equation}\label{glob_loc_subst}
 \alpha \leftrightarrow (2^{-s}z_2, 3^{-s}z_3,5^{-s}z_5,\ldots),
\end{equation}
with the product of scalars replaced by the hermitian product of vectors.

To summarize, we emphasize the contrast between (\ref{QtNCSnonh}) and (\ref{Q_NCS_global}).  It stems from the fact that while $n \mapsto \exp{i Q(n) t}$ is a multiplicative function, 
$n \mapsto \exp{i \Omega(n)^2 t}$ is not. In the latter case, the evolution of the Q-function is not separable, i.e., the phenomenon observed at a single site~$p$ depends on the dynamics observed at other sites. On the other hand, this type of evolution is akin to the classical case with the parameter $\alpha$ determined from the totality of local variables.   We will see in the next section that the similarity extends even further.

\subsection{The Yurke-Stoler interference fringes in the case of NCS} \label{Sec_interfere}

\begin{figure}[t]
\centering
\includegraphics[width=90mm]{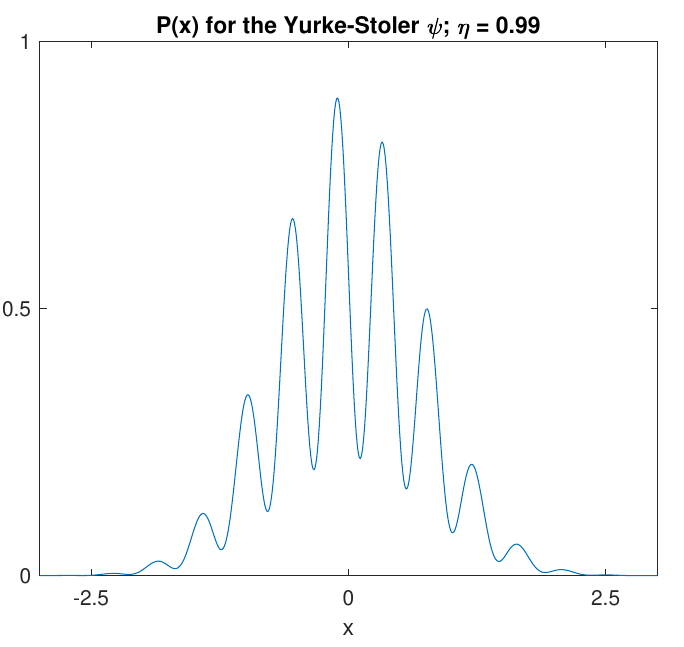}
\caption{The interference pattern for $P(x)$ based on $\psi$ as in (\ref{psip}), $|\alpha| = 5$ and $\eta = .99$. The NCS case driven by the globally quadratic Hamiltonian is completely analogous and results in the same pattern detected via a local observation at any single site $p$ in the infinite array, where $\alpha = p^{-s} z_p$.  }
\label{Numer_YS}
\end{figure}

Yurke and Stoler consider the macroscopically distinguishable superposition (\ref{alpha_doubling}) conjointly with a  beam splitter experiment~\cite{YS86}.
When $\ket{\alpha}$ enters the splitter, the output is  $\ket{\eta^{1/2}\alpha}$ at port A and $\ket{-(1-\eta)^{1/2}\alpha}$ at port B, where a fraction $\eta$ characterizes the splitter's efficiency. More precisely:
\begin{eqnarray}\label{beamsp1}
  \nonumber \ket{\mbox{in}, t} &=&  \ket{\alpha, t}_A \ket{0}_B\\
   &&  \\
  \nonumber \ket{\mbox{out}, t} &=& \ket{\eta^{1/2}\alpha, t}_A \ket{-(1-\eta)^{-1/2}\alpha, t}_B
\end{eqnarray}
Therefore, Eq.~(\ref{alpha_doubling}) shows that,
at time $t+\pi/2$,
\begin{eqnarray}
  \nonumber \ket{\mbox{in}, t+\pi/2} &=&  \ket{\alpha, t+\pi/2}_A \ket{0}_B\\
   &&  \\
  \nonumber \ket{\mbox{out}, t+\pi/2} &=& \frac{1}{\sqrt{2}}\left[e^{i\pi/4}\ket{\eta^{1/2}\alpha, t}_A \ket{-(1-\eta)^{-1/2}\alpha, t}_B
  + e^{-i\pi/4}\ket{-\eta^{1/2}\alpha, t}_A \ket{(1-\eta)^{-1/2}\alpha, t}_B
  \right]
\end{eqnarray}
The phenomenon is then interpreted in the homodyne variable, in which a single CS is expressed by a Gaussian (\ref{psifinal}).
In the homodyne variables, $x$ for port A and $y$ for port B, this output state can be expressed as:
\begin{equation}\label{splt_homodyne}
  \psi_{\mbox{out}} (x,y) = \frac{1}{\sqrt{2}}\left[e^{i\pi/4} \psi_{\eta^{1/2}\alpha}(x)\, \psi_{-(1-\eta)^{-1/2}\alpha}(y)
  + e^{-i\pi/4}\psi_{-\eta^{1/2}\alpha}(x)\, \psi_{(1-\eta)^{-1/2}\alpha}(y)
  \right]
\end{equation}
The interesting object is the probability current through port A, i.e.,
\begin{equation}\label{P_of_x}
  P(x) = \int_{-\infty}^{\infty} dy\,\,  \psi_{\mbox{out}}^* (x,y) \, \psi_{\mbox{out}} (x,y)
\end{equation}
For $\eta$ close to $1$, this probability distribution displays interference patterns, see Fig.\ref{Numer_YS}.

Now, consider the NCS analogue, i.e., the macroscopically distinguishable superposition (\ref{sz_doubling_glob}). Suppose that a beam splitter is placed at one site with the index, say, $q$. To discern the sites, we will denote $\vec{z} = (z_2, z_3, \ldots, z_q, \ldots)$. In analogy with (\ref{beamsp1}), we have
\begin{eqnarray}\label{beamsp2}
  \nonumber \ket{\mbox{in}, t} &=&  \ket{s, (z_2, z_3, \ldots, z_q, \ldots), t}_A \ket{0}_B\\
   &&  \\
  \nonumber \ket{\mbox{out}, t} &=& \ket{s, (z_2, z_3, \ldots, \eta^{1/2}z_q, \ldots), t}_A \ket{s, (z_2, z_3, \ldots, -(1-\eta)^{1/2}z_q, \ldots), t}_B
\end{eqnarray}
Therefore,  at time $t+\pi/2$:
\begin{eqnarray}
  \nonumber \ket{\mbox{in}, t+\pi/2} &=& \left[ \frac{e^{i\pi/4}}{\sqrt{2}} \, \ket{s, (z_2, z_3, \ldots, z_q, \ldots), t }_A + \frac{e^{-i\pi/4}}{\sqrt{2}} \, \ket{s, (-z_2, -z_3, \ldots, -z_q, \ldots), t}_A\right] \ket{0}_B\\
   &&  \\
  \nonumber \ket{\mbox{out}, t+\pi/2} &=& \frac{e^{i\pi/4}}{\sqrt{2}}\ket{s, (z_2, z_3, \ldots, \eta^{1/2}z_q, \ldots), t}_A \ket{s, (z_2, z_3, \ldots, -(1-\eta)^{1/2}z_q, \ldots), t}_B \\
  &+& \frac{e^{-i\pi/4}}{\sqrt{2}}
  \ket{s, (z_2, z_3, \ldots, -\eta^{1/2}z_q, \ldots), t}_A \ket{s, (z_2, z_3, \ldots, (1-\eta)^{1/2}z_q, \ldots), t}_B
  \end{eqnarray}
To find the probability current $P(x_q)$ through port A at site $q$, we need to evaluate an integral analogous to (\ref{P_of_x}). 
However, (\ref{psipcvecz}) indicates that $P(x_q)$ is given by the same formula (\ref{P_of_x}) with the input (\ref{splt_homodyne}), provided $\alpha = q^{-s}z_q$.  
In summary, although the cat-state creation phenomenon is nonlocal in this case, it appears indistinguishable from a local phenomenon when observed locally.

\section{The cat-state creation in the context of generalized bosons}\label{Section_gen_bosons}

We will demonstrate that both the coherent states and the cat-state creation can be generated in the framework of generalized bosons. However, the two phenomena each require a differently constructed state. The coherent states are akin to the celebrated Riemann zeta function. On the other hand, the cat-state creation type, which we call M\"{o}bius states, are tied to the reciprocal of zeta. 
  
Throughout, we adhere to the notation and structures introduced in Subsection \ref{Section_NT}. 
The generalized bosonic creation and annihilation operators act in the same Fock space as their bosonic counterparts. They are defined as follows: 
\begin{equation}\label{bs}
    \hat{b}_p \, |n\rangle = |n/p\rangle ,\quad
       \hat{b}_p^\dagger \, |n\rangle = |np\rangle \quad \mbox{ for any prime } p = 2,3,5, \ldots.
\end{equation}
where, as before, $|n/p\rangle$ is replaced by $0$ in those cases when $p$ is not a divisor of $n$.  
 The following relations are verified via direct calculations (here, $p,q$ stand for arbitrary primes):
\begin{equation}\label{bCCR}
\begin{aligned}
\relax
[\, \hat{b}_p, \hat{b}_q \,] &= 0, \\
[\, \hat{b}_p, \hat{b}_q^\dagger \,] &= \delta_{pq} \, \pi_p, \quad &\text{where } \,\,\pi_p |n\rangle = 
\begin{cases}
0, & \text{if } p \mid n, \\
|n\rangle, & \text{otherwise}.
\end{cases}
\end{aligned}
\end{equation}
Note that $\pi_p$ is the orthogonal projection from the Fock space onto the subspace $\mbox{ span } \{|n\rangle: p \mbox{ is not a divisor of } n\}$. 
It is also natural to extend the definition by setting  
\begin{equation}\label{bms}
  \hat{b}_m = \prod\limits_p (\hat{b}_p)^{a_p(m)} \quad \mbox{ for all integers }  m >1.
\end{equation}
Since all $\hat{b}_p$ commute, $\hat{b}_m$ are well defined. 
In other words, $ \hat{b}_m \, \ket{k} = \ket{k/m}$, and $\hat{b}_m^\dagger \, \ket{k} = \ket{km}$.
It is easily seen that all these operators are bounded, in fact, 
 $  \| \hat{b}_m \| = 
      \| \hat{b}_m^\dagger \| = 1$ for all $m$.
This stands in contrast to the bosonic creation and annihilation operators, which are unbounded.

The corresponding number operators, $\hat{M}_p = \hat{b}_p^\dagger \hat{b}_p$, and $\hat{M} = \sum_{p} \hat{M}_p$ are characterized as follows:
\begin{equation}\label{M_number}
\hat{M}_p \, |n\rangle = \left\{\begin{array}{ll}
\ket{n} & \mbox{ if } p \mbox{ is a divisor of } n \\
\, 0& \mbox{ otherwise } 
\end{array}\right. , \quad \mbox{ and } \quad 
\hat{M} \ket{n} = \omega(n) \, \ket{n},     \end{equation} 
where~$\omega(n)$ is the number of distinct prime divisors of $n$. Note that $\omega(n)$ signifies the number of sites in the support of state $\ket{n}$. Note also that for a state $\ket{\psi} = \sum_n x_n \ket{n}$, we have 
\begin{equation}\label{N_M_expect}
  \langle \psi | \hat{N} |\psi\rangle = \sum_{n} \Omega(n)\,|x_n|^2, \quad 
  \langle \psi | \hat{M} |\psi\rangle = \sum_{n} \omega(n)\,|x_n|^2.
\end{equation}
Thus, while $\hat{N}$ measures the expected number of particles, $\hat{M}$ measures the expected number of occupied sites. 

\subsection{Coherent states for generalized bosons}\label{Nonolal_generalized_CS}

This framework entails an adapted notion of nonlocal coherent states, analogous to (\ref{sz}). We alter notation to make a distinction between the two cases: namely, we replace the sequence $\vec{z}$  by $\vec{\zeta} = (\zeta_2, \zeta_3, \zeta_5, \ldots )\in \mathbb{C}^\omega$. The adapted nonlocal coherent states are defined via:
\begin{equation}\label{sz_gen}
  \ket{ s, \vec{\zeta}\, }   =  \sqrt{S(\vec{\zeta}) } \, \sum_{n=1}^\infty n^{-s}\, \prod \zeta_p^{a_p(n)}\, |\,n\,\rangle,
  \quad \mbox{ where }\quad
S(\vec{\zeta}) = \prod_{p} (1- p^{-2\sigma} |\zeta_p|^2).
\end{equation}
Here, again, $s = \sigma +it$ is fixed and $\sigma>1/2$. The requirement that the normalization factor $S(\vec{\zeta}) $ be well defined and finite imposes a constraint on $\vec{\zeta}$. Note however that finiteness of $P(2\sigma, |\vec{\zeta}|^2)$, defined in (\ref{sz}), implies convergence of the infinite product to a finite limit.  We also assume that $|\zeta_p|^2 \neq p^{2\sigma} $ for all $p$, so that $S(\vec{\zeta}) \neq 0$. Furthermore, the states defined via (\ref{sz_gen}) are product states, i.e. 
\begin{equation}\label{inf_tens_pr_gen}
|\, s, \vec{\zeta}\, \rangle =  \bigodot\limits_{p} \, \ket{p^{-s}\zeta_p}, \quad  \mbox{ where }\,\,  \ket{p^{-s}\zeta_p}= \sqrt{1- p^{-2\sigma} |\zeta_p|^2}\,\sum_{k=0}^{\infty} \, (p^{-s}\zeta_p)^k \, \, |p^k\rangle .
\end{equation}
Note that 
\begin{equation}\label{single_site_gen}
  \hat{b}_p \ket{p^{-s}\zeta_p} = p^{-s}\zeta_p \, \ket{p^{-s}\zeta_p},
\end{equation}
 i.e., states $\ket{p^{-s}\zeta_p}$ are single-site coherent states adapted to the generalized boson framework. Moreover, 
\begin{equation}\label{CS_eq}
  \hat{b}_n \,   |\, s, \vec{\zeta}\, \rangle =  n^{-s}\, \prod \zeta_p^{a_p(n)}\,\, |\, s, \vec{\zeta}\, \rangle .
\end{equation}
In other words, the states $ |\, s, \vec{\zeta}\, \rangle$ are eigenstates of all generalized bosonic annihilation operators simultaneously. They are the nonlocal coherent states adapted to the generalized boson framework.

\subsection{The M\"{o}bius states and the cat-state creation phenomenon} \label{subsection_doubling_gen_boson}

In analogy to the discussion in Subsection \ref{subsection_doubling_NCS}, we examine dynamics driven by the Hamiltonian $\hat{M}^2$, i.e. the square of the generalized number operator. A calculation shows that the states (\ref{sz_gen}) do not exhibit a cat-state creation analogous to (\ref{sz_doubling_glob}). However, there is a family of special states which do have this property. We have dubbed them the M\"{o}bius states. 

Let $\mu: \mathbb{N}\rightarrow \{-1,0,1\}$ be the M\"{o}bius function, i.e., $\mu(1)=1$, $\mu(n)=0$ for those $n$ that are divisible by the square of a prime, and otherwise $\mu(n) = (-1)^{\omega(n)}$. We define M\"{o}bius states as follows 
\begin{equation}\label{inv_sz_gen}
  \ket{ s, \vec{\zeta}\, }_\mu   =  \sqrt{S_\mu(\vec{\zeta}) } \, \sum_{n=1}^\infty \mu(n)\, n^{-s}\, \prod \zeta_p^{a_p(n)}\, |\,n\,\rangle,
  \quad \mbox{ where }\quad
S_\mu(\vec{\zeta}) = \prod_{p} \frac{1}{1+ p^{-2\sigma} |\zeta_p|^2}.
\end{equation}
(Note that, again, finiteness of $P(2\sigma, |\vec{\zeta}|^2)$ implies convergence of the infinite product $S_\mu(\vec{\zeta}) $ to a finite positive number.)
These states are also of the product type, namely:
\begin{equation}\label{Mob_tens_pr_gen}
|\, s, \vec{\zeta}\, \rangle_\mu =  \bigodot\limits_{p} \, \frac{1}{\sqrt{1+ p^{-2\sigma} |\zeta_p|^2}}\, \left( \ket{1} -  p^{-s}\zeta_p \, \ket{p}\right) .
\end{equation}
This formula makes it evident that M\"{o}bius states are not eigenstates of the annihilation operator, as indeed, 
\[
\hat{b}_p \, \left( \ket{1} -  p^{-s}\zeta_p \, \ket{p}\right) = -p^{-s}\zeta_p \ket{1}.
\] 
Next, we introduce short-hand notation: 
\begin{equation}\label{dyn_gen}
   \ket{s, \vec{\zeta}, t}_\mu : = e^{i\hat{M}^2 t} \, |\, s, \vec{\zeta}\, \rangle_\mu .
\end{equation} 
It follows from (\ref{M_number}) that 
\begin{equation}\label{NCSflow_nhm_glob_gen}
 \ket{s, \vec{\zeta}, t}_\mu : =
S_\mu({\vec{\zeta}}) \,\sum_{n=1}^{\infty} \, \mu(n)\, n^{-s} \prod_{p} \zeta_p^{a_p(n)} e^{i\, \omega(n)^2 t}\, \, \ket{n}.
\end{equation}
The following key observations are completely analogous to (\ref{eiksq2}), (\ref{products}), and (\ref{sz_doubling_glob}). First, 
\begin{equation}\label{eiksq2_gen}
  e^{i\omega(n)^2\pi/2}= \frac{1}{\sqrt{2}}\left( e^{i\pi/4} +(-1)^{\omega(n)} e^{-i\pi/4}\right).
  \end{equation}
Second, the special properties of the M\"{o}bius function ensure that, for all $n$, 
\begin{equation}\label{products_gen}
  \prod_{p} \mu(n)\, \zeta_p^{a_p(n)} (-1)^{\omega(n)} = \prod_{p} \mu(n) (-\zeta_p)^{a_p(n)}.
\end{equation}   
From this, it follows that
\begin{equation}\label{sz_doubling_glob_gen}
\ket{s, \vec{\zeta}, t+ \pi/2}_\mu = \frac{e^{i\pi/4}}{\sqrt{2}} \, \ket{s, \vec{\zeta}, t }_\mu + \frac{e^{-i\pi/4}}{\sqrt{2}} \, \ket{s, -\vec{\zeta}, t}_\mu. 
\end{equation}
This is the cat-state creation phenomenon in the generalized boson framework. Note also that we have 
\begin{equation}\label{NCSflow_nhm_glob_gen_minus}
 \ket{s, -\vec{\zeta}, t}_\mu : =
S_\mu({\vec{\zeta}}) \,\sum_{n=1}^{\infty} \, \mu(n)^2\, n^{-s} \prod_{p} \zeta_p^{a_p(n)} e^{i\, \omega(n)^2 t}\, \, \ket{n}.
\end{equation}
Comparing this with (\ref{NCSflow_nhm_glob_gen}) we see that the amplitudes of $\ket{s, \vec{\zeta}, t }_\mu$ and $\ket{s, -\vec{\zeta}, t }_\mu$ at $\ket{n}$ are different whenever $\mu(n) = -1$.  

The problem of physical macroscopic distinguishability of the two states $\ket{s, \vec{\zeta}, t }_\mu$ and $\ket{s, -\vec{\zeta}, t }_\mu$ will not be addressed here. Indeed, at this stage little can be said about methods of generation or detection of M\"{o}bius states. In fact, nothing is known about the ontological status of the specific generalized boson framework discussed here. 
\vspace{.5cm}

\noindent

\emph{Remark.} The cat-state creation in the generalized boson setting has a single site manifestation as well. Note that, according to (\ref{Mob_tens_pr_gen}), the single-site M\"{o}bius state assumes the form 
\[
\ket{\alpha}_\mu := \frac{1}{\sqrt{1+ |\alpha|^2}}\, \left( \ket{1} -  \alpha \, \ket{p}\right),
\]
where we have simplified notation be setting $\alpha = p^{-s} \zeta_p$. Denoting 
$ \ket{\alpha, t} : = e^{i\hat{M}_p^2 t} \, \ket{\alpha}$, we readily obtain 
\begin{equation}\label{single_Mob_doubling}
   \ket{\alpha, t + \pi/2}_\mu = \frac{e^{i\pi/4}}{\sqrt{2}} \, \ket{\alpha, t }_\mu + \frac{e^{-i\pi/4}}{\sqrt{2}} \, \ket{-\alpha, t}_\mu. 
\end{equation}
This is completely analogous with (\ref{alpha_doubling}).

\bibliography{macrosuper}
\end{document}